\documentclass[amssymb,twocolumn,floats,amsmath,superscriptaddress,aps,prb]{revtex4-1}

\usepackage[english]{babel}
\usepackage{natbib}
\usepackage[utf8]{inputenc}
\usepackage[colorinlistoftodos, color=green!40, prependcaption]{todonotes}
\usepackage{amsthm}
\usepackage{mathtools}
\usepackage{physics}
\usepackage{xcolor}
\usepackage{graphicx}
\usepackage[left=23mm,right=13mm,top=35mm,columnsep=15pt]{geometry} 
\usepackage{adjustbox}
\usepackage{placeins}
\usepackage[T1]{fontenc}
\usepackage{csquotes}
\usepackage[hidelinks]{hyperref}
\usepackage{xcolor}
\hypersetup{
    colorlinks,
    linkcolor={red!50!black},
    citecolor={blue!50!black},
    urlcolor={blue!80!black}
}

\def \UR{Institute of Theoretical Physics, University of Regensburg, 93040 Regensburg, Germany}
\def \IFT{Institute of Theoretical Physics, Faculty of Physics, University of Warsaw, Pasteura 5, 02-093 Warsaw, Poland}
\def \PW{Department of Semiconductor Materials Engineering Faculty of Fundamental Problems of Technology, Wrocław University of Science and Technology Wybrzeże Wyspiańskiego 27, 50-370 Wrocław, Poland}

\begin{document}

\title{Magneto-optical anisotropies of 2D antiferromagnetic MPX$_3$ from first principles} 

\author{Miłosz Rybak}\affiliation{\PW}
\author{Paulo E. {Faria~Junior}}\affiliation{\UR}
\author{Tomasz Wozniak}\affiliation{\PW}
\author{Pawel Scharoch}\affiliation{\PW}
\author{Jaroslav Fabian}\affiliation{\UR}
\author{Magdalena Birowska}\affiliation{\IFT}
\email[Correspondence email address: ]{Magdalena.Birowska@fuw.edu.pl}

\date{\today} 

\begin{abstract}
Here we systematically investigate the impact of the spin direction on the electronic and optical properties of transition metal phosphorus trichalcogenides (MPX$_3$, M=Mn, Ni, Fe; X=S, Se) exhibiting various antiferromagnetic arrangement within the 2D limit. Our analysis  based on the density functional theory and versatile formalism of Bethe-Salpeter equation reveals larger exciton binding energies for MPS$_3$ (up to 1.1 eV in air) than MPSe$_3$ (up to 0.8 eV in air), exceeding the values of transition metal dichalcogenides (TMDs). For the (Mn,Fe)PX$_3$ we determine the optically active band edge transitions, revealing that they are sensitive to in-plane magnetic order, irrespective of the type of chalcogen atom. We predict the anistropic effective masses and the type of linear polarization as an important fingerprints for sensing the type of magnetic AFM arrangements. Furthermore, we identify the spin-orientation-dependent features such as the valley splitting, the effective mass of holes, and the exciton binding energy. In particular, we demonstrate that for MnPX$_3$ (X=S, Se) a pair of non equivalent K+ and K- points exists yielding the valley splittings that strongly depend on the direction of AFM aligned  spins. Notably, for the out-of-plane direction of spins, two distinct peaks are expected to be visible below the absorption onset, whereas one peak should emerge for the in-plane configuration of spins. These spin-dependent features provide an insight into spin flop transitions  of 2D materials. Finally, we propose a strategy how the spin valley polarization can be realized in 2D AFM within honeycomb lattice.

\end{abstract}

\keywords{first keyword, second keyword, third keyword}

\maketitle

\section{Introduction}
The subtle interplay between the spin, charge, orbital, and lattice degrees of freedom driven by the electron correlation is one of the key aspects in condensed matter physics behind novel electronic phases of matter and intriguing physical phenomena. In particular, the electronic properties can be modified whenever the spin direction is altered, as spin-orbit coupling (SOC) depends on the spin direction. For instance, spin-valley coupling serves as the fundamental mechanism in optically-controlled valley polarization \cite{Zeng2012}, spin-Hall and valley-Hall effects \cite{PhysRevLett.108.196802}. Contrast to most of the findings in non-magnetic 2D materials \cite{Tombros2007, Erve2012, Kamalakar2015}, where pseudo-spins are involved, 2D magnets exhibit active carrier spins,  enabling  studies of magnetism in reduced dimensions \cite{Gibertini2019}.
 
Unlike ferromagnets (FMs), the antiferromagnets (AFMs)  are commonly found in nature and they are permitted in each
magnetic symmetry group, however they are less utilized than FMs \cite{Nemec2018}. Currently, AFM materials are considered as promising candidates for future spintronic applications, due to unique properties including  insensitivity  to external magnetic fields,  lack of stray fields, and ultrafast spin dynamics in the terahertz regime \cite{InfoMat}.  The spontaneously long-range ordered of microscopic magnetic  moments, resulting in zero net magnetization, which  makes the AFMs  insensitive to external magnetic fields. Thus, a control of AFM state requires  very high magnetic fields  and demands  unconventional means of  detection \cite{Nemec2018}.

Many properties such as optical, electronic, and vibrational, rely on the magnetic  ordering\cite{BirowskaPRB, Autieri2022, D3TC01216F}, as well as magnetic moment orientation\cite{Liu2019}.  In particular, a giant impact of the spin direction on the band structure, have been recently demonstrated for 2D ferromagnetic CrI$_3$ material \cite{doi:10.1021/acs.nanolett.8b01125}. Although, the spin-direction properties are reported for conventional and layered FMs \cite{Sander2004, 10.1063/1.1381100}, the research on 2D AFM materials is very limited and scarce \cite{Nemec2018, Gibertini2019, 7109970, Rahman2021}.



One of essential feature for manipulating AFM state is the  magnetocrystalline anisotropy (MAE) which is evident in spin flip or spin flop transitions. The latter one requires relatively weak MAE and it is reported in literature for AFM compounds exhibiting various electronic states such as topological insulators \cite{PhysRevLett.125.037201}, layered materials \cite{PhysRevB.90.104412}, and conventional semiconductors \cite{PhysRevB.95.104418}. In addition, the experimental techniques including indirect means of magnetic phases such as anisotropic magnetoresistance (AMR), anomalous Hall effect (AHE) or second harmonic generation (SGH)\cite{Jungwirth2016, Nemec2018,Fina2014}, are employed to detect the AFM orientation in 2D materials. Interestingly, the magnetic phase transition of controlled anisotropic phenomenon in layered magnets provides crucial understanding of fundamental magnetism in reduced dimensions. 
 
The present work is motivated by the lack of systematic studies regarding the impact of AFM orientation of magnetic moments on the optoelectronic properties of 2D materials. Hence, in this paper we put particular attention to pinpoint the magnetic fingerprints in indirect properties that can engineer the AFM ordering.
We conduct a theoretical analysis based on density functional theory (DFT) and effective Bethe-Salpeter equation (BSE) to identify various spin direction features, which can be utilised in spin-processing functionalities. This study focuses on spin angle evolution of the effective masses of carriers,  optically active band edge transitions, exciton binding energies, which remain almost unexplored in the context of van der Waals (vdW) AFM crystals. 

Here, we present a systematic study of the impact of spin orientation on the electronic and optical properties of series of the monolayers of MPX$_3$, where M= Mn, Ni, Fe,  and X= S, Se, assuming collinear arrangement of the magnetic moments. The results are presented as following: first we examine the magnetic ground state, determining easy and hard axes of magnetization at the level of PBE+U+SOC approach. Next, we consider the electronic features such as band extrema, effective masses, valley splitting and excitonic properties including the excitonic binding energy. Although, the electronic properties  have been widely reported for particular spin arrangement and direction, the impact of the  orientation of the collinearly ordered spins on opto-electronic properties MPX$_3$  are largely missing. Finally,  band edge excitons have been  systematically predicted for this class of materials for the first time. 





\section{Computational Details}

The calculations were performed in the framework of density functional theory (DFT) using the generalized gradient approximation within the PBE  flavor \cite{doi:10.1063/1.1926272}, as implemented in VASP software \cite{KRESSE199615}. The ion–electron interactions were described by the projector augmented wave (PAW) method  \cite{Holzwarth2001}. Plane-wave basis cutoff and $\Gamma$ centered Monkhorst-Pack \cite{Monkhorst1976} k-point grid were set to 500 eV and $10\times6\times1$, respectively. A Gaussian smearing of 0.05 eV was employed for the Brillouin zone (BZ) integration. The interlayer vdW forces were treated within Grimme scheme using D3 correction \cite{doi:10.1063/1.3382344}. A vacuum of the thickness equal to 20 \AA ~ was added to mimic the isolated monolayer. Most of the results were obtained  using PBE+U method based on Dudarev's approach, with the effective on-site Hubbard U parameter (U$_{eff}$ =U-J, where J=1eV) assumed for $3d$ orbitals. Note that U$_{eff}$ is hereafter denoted as U. To check its impact on various properties two values of U= 3, 5 eV were employed. The SOC within the non-collinear treatment of magnetism  was taken into account on the top of the PBE+U scheme. The position of the atoms and unit cell were fully optimized within the PBE+U approach.  In order to predict the magnetic easy and hard axis, the spins of the magnetic ions have been rotated from out-of plane to in-plane directions with polar angle step $\Delta\Theta$=15$^0$. For every $\Theta$ the atomic positions and the lattice parameters have been fully optimized.
The effective masses of carriers at the band edges  were examined using finite difference method \cite{fonari2015effective}.
The direct interband momentum matrix elements were computed from the wave function derivatives using Density Functional Perturbation Theory (DFPT)\cite{PhysRevB.73.045112} in order to determine  the optically active transitions, as discussed in  \cite{PhysRevB.101.235408}.  The macroscopic optical dielectric constants frequency $\epsilon_{\infty}$ were obtained using DFTP  in the independent particle approach with (DFT-TD) and without local field effects (IP) \cite{PhysRevB.73.045112}

Exciton calculations were performed within the effective BSE formalism\cite{PhysRevLett.80.4510, Rohlfing2000PRB, Zollner2019PRB, Zollner2020PRB, Zollner2023PRB, BirowskaPRB} using the effective masses and dielectric constants obtained from the DFT calculations. We solve the effective BSE numerically using the parameters given in Ref.\cite{BirowskaPRB}.

\begin{figure*}[ht]
    \centering
    \includegraphics[scale=.18]{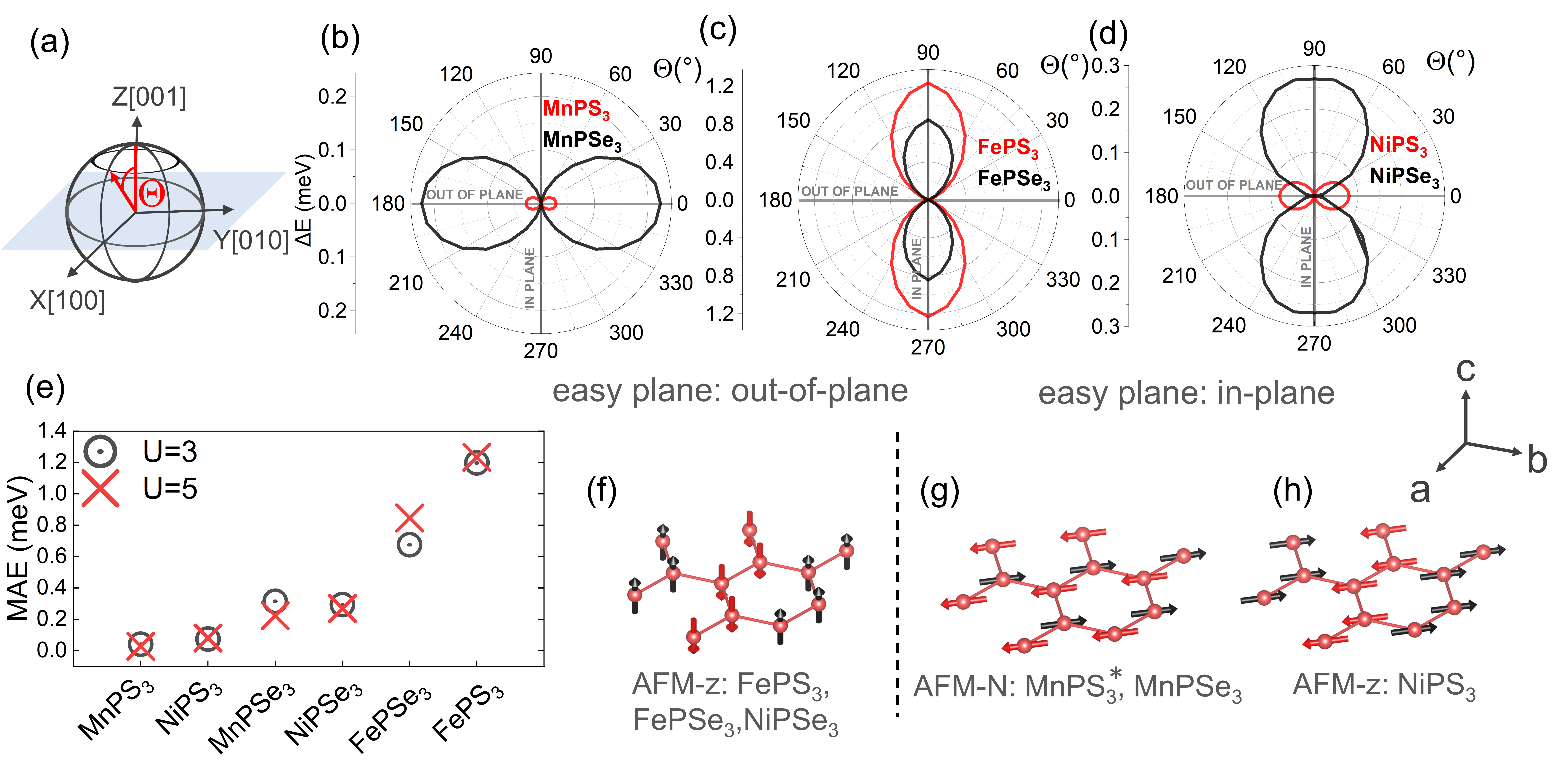}
    \vspace{0pt}
    \caption{\label{spins} (a) The rotation angle $\Theta$  of collinearly aligned AFM spins directed towards the (001) plane. (b-d) Polar plots of energy  difference $\Delta E$ between the  particular direction of the spins and the lowest energy configuration  for Hubbard U=5 eV. (e) The magnetocrystalline energy (MAE) is defined here, as a difference between the energies of spins parallel to hard (highest energy) and easy axes (lowest energy) of magnetization. (f-h) Structural arrangement of the spins exhibited by the magnetic ground state for all employed systems. The direction of the spins are parallel to the easy axis / plane of magnetization. The $\Delta E$ and the MAE are given in meV per magnetic ion.}
\end{figure*}
\section{Results}



\subsection{Magnetic ground state $\&$ magnetic easy axis}

First, we examine the impact of the rotation of spins on the energy profile  to determine the magnetic ground state and magnetic easy/hard axes for all employed MPX$_3$ systems. We assume a collinear alignment of the spins and angle of rotation $\Theta$ (see Fig. \ref{spins} (a)) varied from the out-of plane to in-plane configuration. The rotation of the spins within the basal plane yields up to two orders of magnitude smaller energy changes compared to out-of plane ones, hence in-plane rotations are not further considered.

The magnetic ground state exhibits antiferromagnetic Neel (AFM-N) and antiferromagnetic  zigzag (AFM-z) arrangement of the spins for Mn and M=Ni, Fe, respectively (see Fig. \ref{spins}), in line with other reports \cite{PhysRevB.91.235425, BirowskaPRB, Autieri2022, Budniak2022}. The MPX$_3$ materials are reported to be robust antiferromagnets \cite{Autieri2022} and even high concentration of the substitutional dopants could not alter the magnetic ordering of the host \cite{PhysRevResearch.4.023256, Autieri2022}.  Additionally, changing the angle of spin alignment $\Theta$ requires at least one order of magnitude lower energy (tens of meV per magnetic ion) than  change of the AFM ordering (at least tens of meV per magnetic ion). The computed energy  difference ($\Delta E$) between the out-of plane and in-plane directions can be one order of magnitude greater for Se than S compounds (see Fig. \ref{spins} (b-d)), which is expected for heavier atoms, exhibiting larger SOC coupling \cite{PhysRevResearch.4.023256}. Surprisingly, FePS$_3$ exhibits a larger MAE than FePSe$_3$ (see Fig \ref{spins}(e)). The value of U does affect neither the type of AFM ordering nor the direction of the magnetic easy axis. However, the $\Delta$E depends on  the Hubbard U parameters. Namely, the smaller values are generally obtained for larger U, except for FePS$_3$ monolayer (see Fig. S2). For the employed materials, an increase in the effective U resulted in larger lattice constants by up to 1.5 \% and larger magnetic moments (see Table S2).


Let us now consider the easy axes of magnetization, which  are presented in Fig.~\ref{spins}(f, g) for all employed structures,  with the hard axes predicted to be orthogonal to the corresponding easy axis/plane. In particular, for FePX$_3$ the magnetic easy axis points along the $c$ crystallographic direction as reported in previous studies \cite{Olsen_2021,PhysRevB.92.224408,19833919,molecules26051410,PhysRevB.91.235425,PhysRevB.91.235425,WIEDENMANN19811067}. Hence, this kind of systems are considered to be Ising-type antiferromagnets, with a strong uniaxial magnetic anisotropy. 

For the  MnPX$_3$ compounds, we predict that the easy plane coincides with monolayer plane, unlike the recent experimental report for MnPS$_3$, which has demonstrated that the spins are slightly tilted from the $c$-axis \cite{PhysRevB.82.100408}. The direction of the ordered spins for MPX$_3$ compounds result from the interplay between the single-ion anisotropy and magnetic dipolar  interactions (MDIs), as already discussed in a recent publication \cite{PhysRevResearch.4.023256}. In addition, the spin-orbit splitting is negligible for MnPS$_3$, which manifests itself as the lowest MAE for all employed structures (see Fig. \ref{spins}(b,e)). Thus, the MDI, generally regarded to be weak\cite{PhysRevB.93.014421}, might be decisive in determining the direction of the MnPS$_3$ spins. Note, that the MDIs are not accounted for in our PBE+U+SOC approach, and will be considered elsewhere. On the other hand, in the case of the MnPSe$_3$, the MDIs could be neglected due to larger lattice parameters and SOC. Hence, the magnetic easy axis of MnPSe$_3$ coincides with the monolayer plane, in agreement with other reports in the literature\cite{Sen_2020, WIEDENMANN19811067, 19833919}, and confirmed by spin flop transition reported upon the nonmagnetic substitution in MnPS$_3$ thin films \cite{PhysRevResearch.4.023256}. Regarding NiPS$_3$, the in-plane position of the magnetic moments are preferable and the magnetic ordering can be described by the XY Heisenberg Hamiltonian\cite{PhysRevB.92.224408, Olsen_2021, Hwangbo2021, Belvin2021}. Our results predict the easy axis of magnetization within the monolayer frame, without any deviation in the z direction as reported recently for monolayer\cite{Kim2019}. However, we do not exclude the possibility that the inclusion of the MDIs interaction can facilitate the rotation of the spins towards the out of the plane direction. For the case of NiPSe$_3$, we determined the easy axis of magnetization to be in out-of-plane direction in line with recent report \cite{SUN2023101188}. Similar results regarding the easy axis/plane of magnetization have been reported for the corresponding bulk systems \cite{PhysRevResearch.4.023256}.



\begin{figure*}[t]
    \centering
\includegraphics[width=1.0\textwidth]{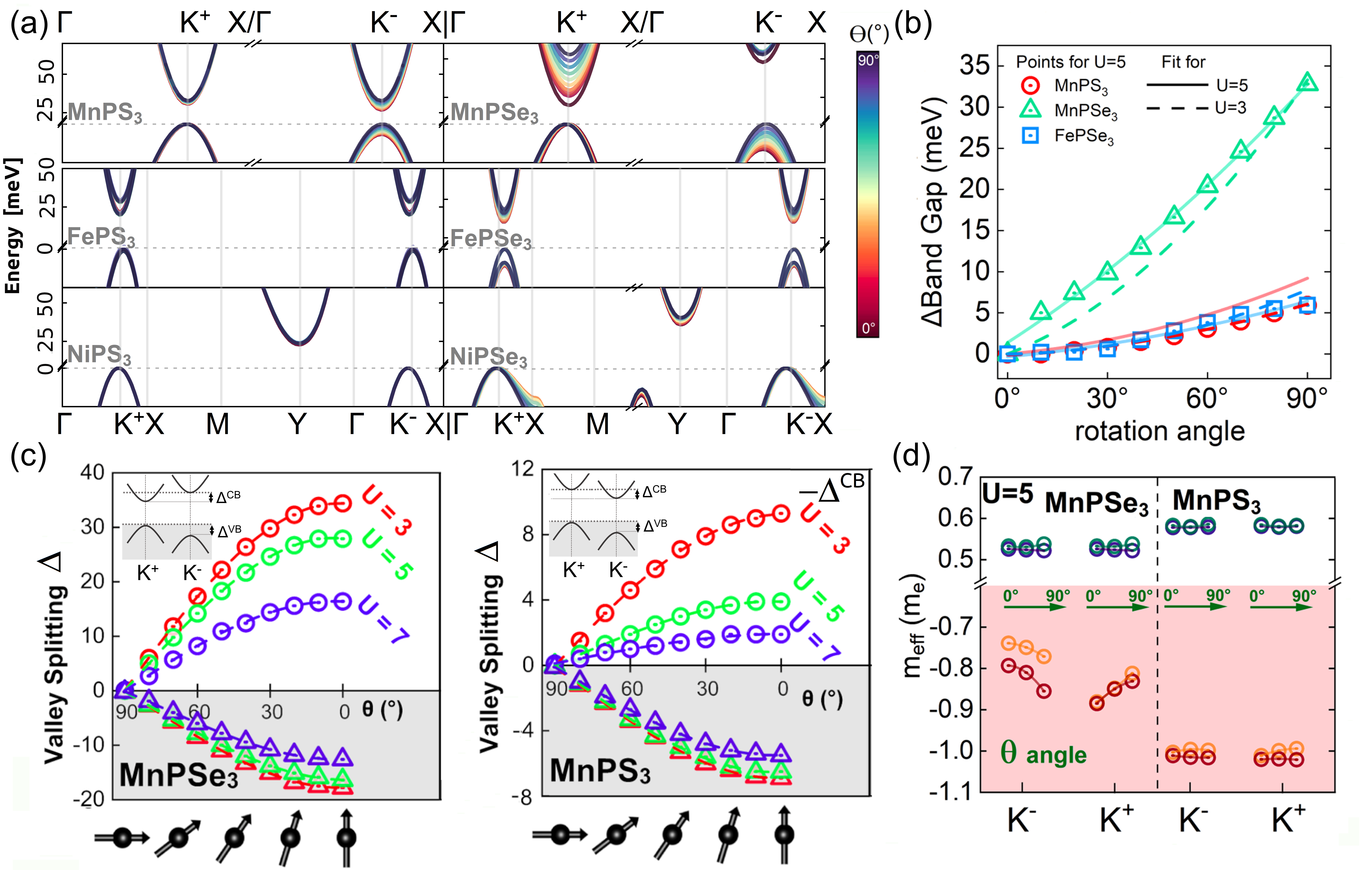}
    \vspace{0pt}
\caption{\label{fig:vs} Spin-orientation electronic properties. (a)Band extrema for all employed monolayers  as a function of spin direction (color scale). (b) Enhancement of the electronic band gaps $\Delta E_{gap}$ for particular spin directions in respect to out-of-plane spins.  (c) Valley splitting $\Delta^{VB}$, $\Delta^{CB}$  are defined $\Delta^{VB}$=E$^{VB}_{K_-}-$E$^{VB}_{K+}$, $\Delta^{CB}$=E$^{CB}_{K_-}-$E$^{CB}_{K+}$, respectively for Mn and Fe contained monolayers, given in meV. Note, that for the clarity of picture regarding MnPS$_3$, the $-\Delta^{CB}$ is plotted. (d) In-plane components of effective mass tensor for electrons (positive masses) and holes (negative masses) of MnPX$_3$ in respect to spin direction and for two non-equivalent $\pm K$ valleys.}
\end{figure*}

\begin{table*}[ht]
\caption{\label{tab:MPX3} Spin-dependent electronic features for direct band edge transition occurring at $\pm K$ valleys (VBM$\rightarrow$CBM) in the presence of SOC (PBE+U(5eV)+SOC approach). The valley splitting  of the VBM ($\Delta^{VB}$) and CBM ($\Delta^{CB}$)  are defined as $\Delta^{VB}$= E$^{VB}_{K_-}-$E$^{VB}_{K+}$ and  $\Delta^{CB}$ = E$^{CB}_{K_-}-$E$^{CB}_{K+}$), respectively. The  spin degeneracy of the band edges $\delta^{VB}$ and $\delta^{CB}$ are calculated as $\delta^{VB}$=| E$_{\uparrow}^{VB}-$E$_{\downarrow}^{VB}|$ and $\delta^{CB}$=| E$_{\uparrow}^{CB}-$E$_{\downarrow}^{CB}|$, respectively. We refer intensity (Inten.) to the oscillator strength of of the band edge  transitions. The optically active transitions coupled to linear polarization of light (x,y,z) are collected in the last column.  The intensity is given by $\left|\frac{\hbar}{m_{e}}\hat{e}\cdot\vec{p}_{cv}\right|^{2}$ in which  $\hat{e}=\{\hat{x},\hat{y},\hat{z}\}$ is the light polarization (pol.) and $\vec{p}_{cv}$ is the matrix element between CBM and VBM. } 
 \def\arraystretch{1.1}
\begin{center}\label{features}
\begin{tabular}{  c|  c| c| c|c |c }
\hline
\hline
  MPX$_3$ (magn. state)& spin direction ($\Theta$)& K+/K- ($\Delta^{VB}$, $\Delta^{CB}$)& Spin Deg. ($\delta^{VB}$, $\delta^{CB}$)& Hex. symm. &(pol.) Inten. [$(\textrm{eV\AA})^{2}$] \\
 \hline
 \textbf{MnPS$_3$} (AFM-N)) & 0$^0$ &  $\checkmark$ (-6.4, -4.4)  &  X & $\checkmark$   &(z) 0.59 \\
& 90$^0$ & X (0, 0) & X & $\checkmark$ &(z) 0.62   \\
  \hline
   \textbf{MnPSe$_3$} (AFM-N)& 0$^0$ &  $\checkmark$ (-17.7, 33.1)  &  X & $\checkmark$   & (z) 0.13  \\
 & 90$^0$ & X (0, 0) & X & $\checkmark$ &(z) 0.18; (x,y) 0.05\\
  \hline
\textbf{FePS$_3$} (AFM-z) & 0$^0$&  X  &  $\checkmark$ (1.8, 9.1)& broken  &  0.71 (y)  \\
& 90$^0$ &   X& $\checkmark$ (1.5, 7.4) & broken &  0.73 (y) \\
  \hline
\textbf{FePSe$_3$} (AFM-z)& 0$^0$&  X  &  $\checkmark$ (7.4, 3.1) & broken  &   (y) 0.27; (x) 0.04 \\
& 90$^0$ &   X& $\checkmark$ (9.3, 5) & broken &   (y) 0.29; (x) 0.02 \\
\hline
\hline
\end{tabular}
\end{center}
\end{table*}
\subsection{Electronic properties}
\subsubsection*{electronic band gaps}
Our results reveal that the structures containing Mn and Fe exhibit direct band gaps located at K high symmetry point (see Fig. \ref{fig:vs} (a)), in line with previous reports \cite{BirowskaPRB,Budniak2022,arxivEllenore}, while the NiPS$_3$ system is an indirect semiconductor confirmed by experiments \cite{Hwangbo2021}.  In particular, for FePS$_3$ and NiPSe$_3$ there is a strong contribution of the  $3d$ states to the bands in vicinity of Fermi level. Hence, for these systems, significant changes of the position of the band extrema (VBM, CBM) and their curvatures  upon the Hubbard U parameter are visible (see Tables S2 and S3 and discussions therein). 

\subsubsection*{U dependence}
For all employed compounds, except MnPX$_3$, the U parameter  significantly affects the curvature of the conduction bands as reported previously \cite{Budniak2022}, and hence the effective masses of holes. Hence, all features are examined assuming two Hubbard U parameters U=3, 5 eV). In particular, the large value of the effective masses of electrons exhibited for Ni contained systems,  indicating a flat character of the conduction bands, which  can be further enhanced by adopting  larger values of U (see Table S3).  By applying the Hubbard U to 3d states, one can shift the 3d states away from the Fermi level. In particular, for MnPX$_3$ structures the conduction bands are mainly build from p-states, and further shifting 3d states do no affect the band curvature nor the band gap value, whereas for FePX$_3$ compounds the conduction bands consist mainly of 3d states, hence further enlargement of U strongly impacts electronic features around band edges. Additionally, our results corroborate the previous reports on strong dependence of Hubbard U on fundamental energy band gap (see Table S2) \cite{Budniak2022}. 
\subsubsection*{Effective masses}
For all of employed monolayers, the  heaviest effective mass is obtained for NiPX$_3$, whereas the lightest one is demonstrated for MnPSe$_3$ (see Table S3 and Fig. S3). Generally, the lower effective masses are exhibited by the electrons than holes, and for Se contained structures  than  corresponding S systems (by about 30$\%$), except for NiPX$_3$ structures where these trends are opposite. Additionally, the anisotropic in-plane components of effective masses (m$_1 \neq$ m$_2$) are exhibited by monolayer with AFM-z ground state, whereas for MnPX$_3$ monolayers with AFM-N ordering, the in-plane components are isotropic (m$_1$=m$_2$). Note, that the AFM-z magnetic arrangement breaks the hexagonal symmetry, as it consist of spins  ferromagnetically aligned along zigzag chain (see Fig 1(g)), whereas the AFM-N type of magnetic order is commensurate with the structural symmetry.  Hence, the anisotropic effective masses, and thus, anisotropic transport properties can be regarded as a magnetic marker   distinguishing different type of  antiferromagnetic ordering within hexagonal lattice. In addition, the inclusion of the SOC does not affect the curvature of examined systems, except for MnPSe$_3$ where slight changes are shown for K+ and K- valleys (see Table S4).


\subsubsection*{Spin-orientation-dependent electronic features in MPX$_3$ (M=Mn, Fe; X=S,Se)}
 Now, let us make  closer inspection on the \textbf{band edges} of MLs exhibiting direct transitions. The electronic features are collected in Table \ref{tab:MPX3}. Note, that in the absence of the SOC, the band extrema are doubly degenerated for all employed systems. The presence of SOC in the structures containing Mn preserves the spin degeneracy of the band extrema (VBM, CBM), however causing the valley splitting at +K and -K, which is expected for the honeycomb lattice with AFM-N arrangements of the spins with SOC included\cite{PNAS}. Namely, the SOC preserves the spin degeneracy but leads to a renormalization of the valleys (+K and -K are not equivalent) for AFM-N magnetic ground state. The size of the valley splitting ($\Delta$) depend on the chalcogen atoms (larger values for Se atoms), Hubbard U, and spin directions (see Fig. \ref{fig:vs} and Tables \ref{features}, S6). In particular, the largest valley splitting is attained for the out-of-plane direction of the spins ($\Delta^{CB}=39$ meV for MnPSe$_3$ and U=3eV)  and lower value of U as presented in Fig. \ref{fig:vs}(c), whereas no polarity of the valleys are observed for the in-plane direction of the spins. Namely, the rotation of the spins towards out-of-plane directions results in enhancement of the valley splitting. Hence, the band gap of MnPSe$_3$ can be changed by up to 35 meV as presented in Fig. \ref{fig:vs}(b, c),  whereas for the rest of the compounds the minor changes are observed  (see Fig. \ref{fig:vs}(b)). Additionally, the effective masses of holes for MnPSe$_3$ are sensitive to the direction of the spins as presented in Fig. \ref{fig:vs}(d).  Interestingly, in the case of MnPS$_3$, the band gap changes its character from direct to indirect when the spins are rotated from an out-of-plane  to an in-plane directions. 
\begin{figure*}[ht]
    \centering
\includegraphics[width=1.0\textwidth]{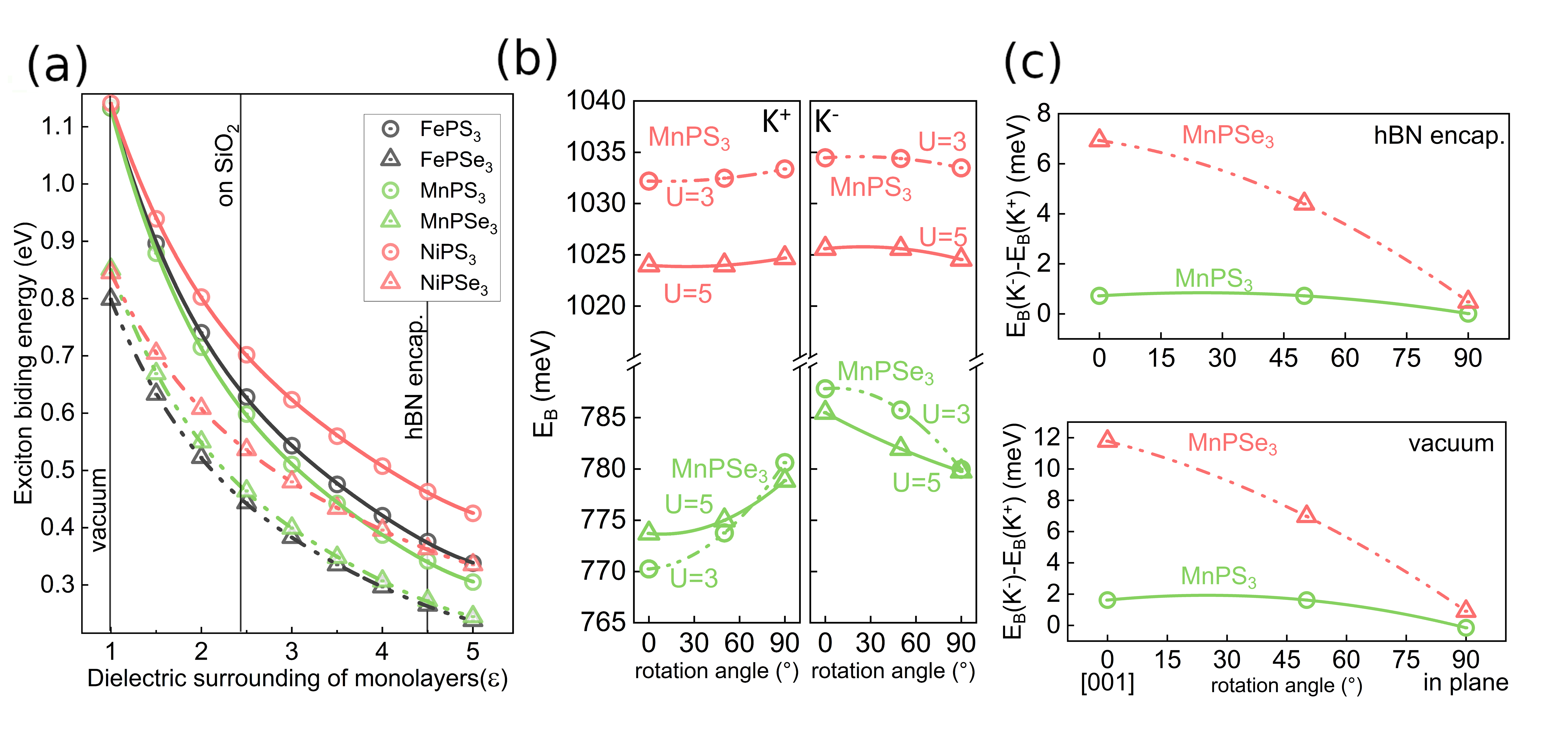} \vspace{0pt}
\caption{\label{fig:excitons} (a) Large   exciton binding energies obtained for all considered monolayers with magnetic ground state with respect to the dielectric screening of the surroundings. The values for air/vacuum, SiO$_2$ \cite{PhysRevB.88.045318} substrate, and hBN encapsulation \cite{Stier2018PRL} are denoted by vertical lines. (b) Evolution of the E$_B$ with respect to the spin direction for two values of Hubbard U (3 and 5 eV). (c) The energy difference  between the E$_B$ calculated for $-K$ and $+K$ valleys as a function of the spin direction angle $\Theta$ (within DFT+U=5eV and  DFT-TD approaches), for hBN encapsulated and freestanding layers.}
\end{figure*}

Regarding the monolayer of FePX$_3$, the presence of SOC and local inversion symmetry breaking in FeX$_6$ octahedra lift the degeneracy of the band extrema (few meV), irrespective to the spin direction (see column 4 in Table \ref{features}). In particular,  larger spin polarization $\delta$ is obtained for CBM than VBM ($\delta^{CB}$= 9.1 meV, $\delta^{VB}$= 1.8 meV), and for higher values of U for FePS$_3$. In contrast to the FePS$_3$, the  opposite trends are obtained for FePSe$_3$ (for the details see Table S6). Hence, the spin splitting of the bands is obtained for  hexagonal lattices exhibiting in-plane structural anisotropy, as reported for MnPS$_3$ (see Fig. 5(d) in \cite{arxivEllenore}). In this case the spin polarization occurred, alongside with renormalization of the +K/-K valleys as expected for AFM-N.  In addition, one of the consequences of the local inversion symmetry breaking exploited in Fe monolayers is a  lattice distortion reflected in the ratio of the lattice parameters deviated from hexagonal symmetry ($b/a\neq\sqrt{3}$) \cite{arxivEllenore, Gosh}. 


\subsection{Excitonic properties}

\subsubsection*{Dielectric properties}
We start our discussion by analyzing the dielectric properties of the employed systems. Generally, the in-plane components of static dielectric constants ($\varepsilon_{xx}$, $\varepsilon_{yy}$)  are isotropic for all studied systems, except for monolayers of (Ni,Fe)PX$_3$, where they differ by up to 4$\%$  (see SI and Tabs. S7, S8). Due to the fact that dielectric tensor is well defined for the bulk materials, we calculate the 2D polarizability $\chi_{\parallel}$ (dielectric screening length, for the details see SM). Our results reveal that the dielectric screening length increases as the atomic number of the metal also increases $\chi_{\parallel}^{Mn} > \chi_{\parallel}^{Fe} > \chi_{\parallel}^{Ni}$, exhibiting the smallest and highest values  equal to 3.25 \textrm{\AA} (MnPSe$_3$, U=5 eV)  and 6.13 \textrm{\AA} (NiPSe$_3$,  U=5eV), respectively. In addition, the systems containing Se exhibit $\sim$40$\%$ larger screening properties than their corresponding S systems. The screening properties of MPX$_3$ compounds are generally smaller than for other vdW structures such as MoSe$_2$ (8.23 \textrm{\AA}). Note that for the (Ni,Fe)PX$_3$ systems, the dielectric screening lengths strongly depend on the Hubbard U. In particular, the larger  values of $\chi_{\parallel}$ are attained for the smaller Hubbard U, which  correlates with the strong impact of the U on the DOS (see Fig. S2 (b)).  

\subsubsection*{Band edge excitons}
Let us now examine the band edge excitons using the relevant information from DFT ($m_{eff}$, $\chi_{\parallel}$) combined with the effective BSE\cite{PhysRevLett.80.4510, Rohlfing2000PRB}. We show the exciton binding energies (E$_B$) in Fig.~\ref{fig:excitons}(a) (for the details of calculations see Ref.\cite{BirowskaPRB}). All employed structures exhibit large E$_B$ exceeding 1 eV and 0.8 eV for bare monolayers of MPS$_3$ and MPSe$_3$, respectively, outperforming the values in TMDCs \cite{BirowskaPRB} (the exact values are collected in Tables S9, S10). The excitons have been recently experimentally reported for few layers of NiPS$_3$ \cite{Kang2020,Hwangbo2021, Wang2021}, and FePS$_3$ \cite{arxivEllenore}, however, their origin is  still under a hot debate. Generally the E$_B$ are larger by about 30$\%$ for S contained structures than corresponding Se compounds, mainly due to larger effective masses and smaller dielectric screening for MPS$_3$. The E$_B$ decreases as the effective dielectric constant of the environment increases, preserving the same trend for all monolayers. This can be explained within the oversimplified exciton picture (hydrogen model) for which E$_B$ is proportional to effective mass and inversely proportional to the square of dielectric screening.

\subsubsection*{Optical transitions and selection rules}
Besides the binding energy, it is also relevant to determine the selection rules of the direct band edge transitions, which are summarized in Table \ref{tab:MPX3}. All of the transition are optically allowed (non-zero oscillator strength) exhibiting linear polarization of light. In particular, for monolayers with AFM-N phase (MnPX$_3$) the polarization of light is along $z$ direction, whereas for the AFM-z (FePX$_3$) pointing along $y$ direction. Interestingly, similar conclusions are observed assuming various AFM metastable phases within the same magnetic material \cite{BirowskaPRB}. Therefore, the polarization of light is sensitive to magnetic order irrespective of the type of transition metal and chalcogen atoms. This is in line with recently reported linear polarization of the
sharp emission, that aligned perpendicular to the spin orientation \cite{Wang2021}.  Hence, the polarization of light might be a tool to distinguish the type of AFM ordering. All of these direct band edge excitons  are optically active transitions and couple to $z$-polarized light. Comparing with the widely studied TMDCs, these transitions in MPX$_3$ systems have an intensity two orders of magnitude smaller than the bright (A and B) transitions but have comparable intensity to the dark (D) transitions\cite{PhysRevB.101.235408, FariaJunior2022NJP}. 

\begin{figure*}[ht]
    \centering
\includegraphics[width=0.8\textwidth]{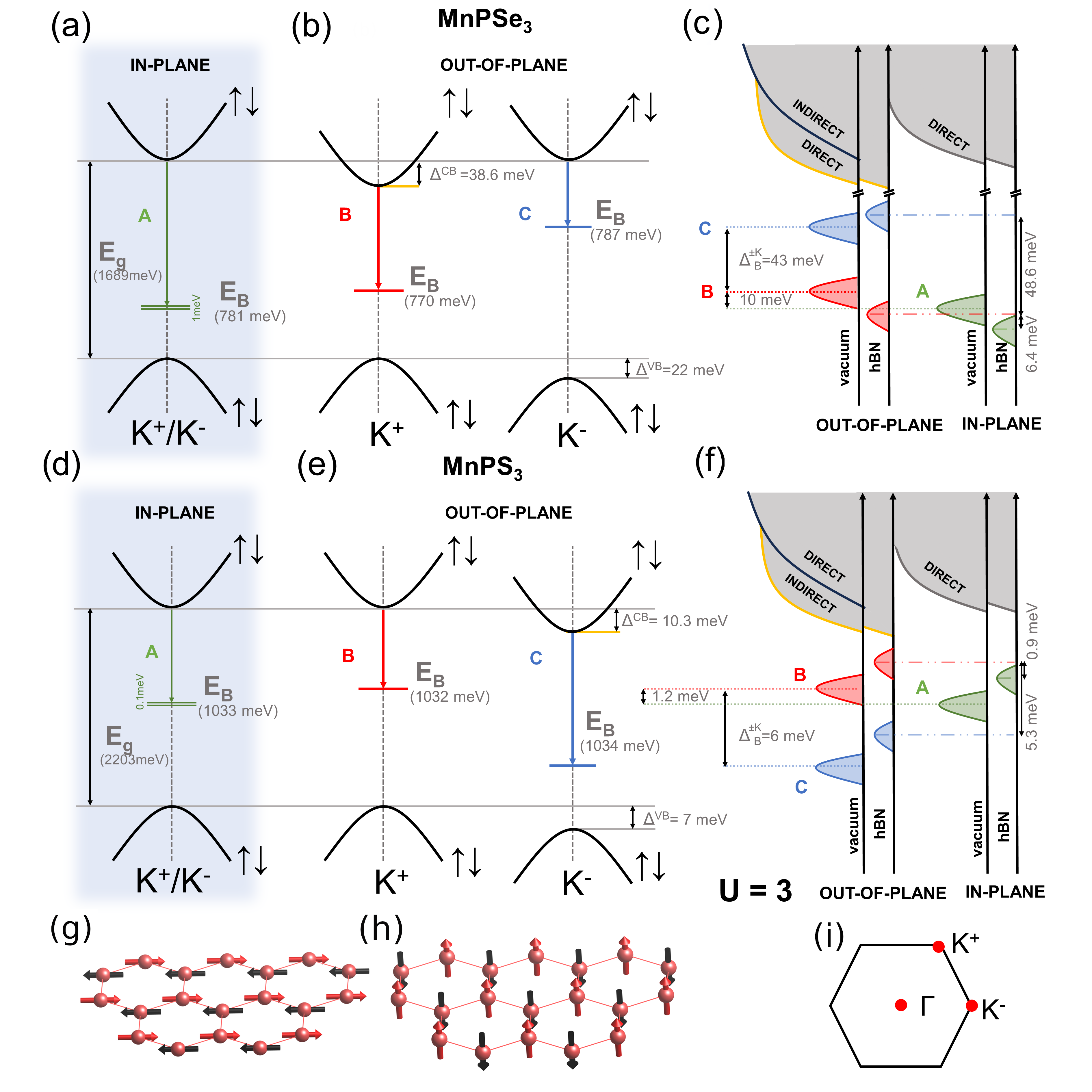}
    \vspace{0pt}    \caption{\label{fig:transitions} Schematic diagrams (U=3 eV) of   direct transitions for (a, b) MnPSe$_3$ and (c, d) MnPS$_3$ exhibiting in-plane and out-of-plane directions of the AFM aligned spins, as schematically denoted  at the bottom of the picture (g, h). Schematic absorption onsets (c, f) with depicted in coloured exciton peaks A,B and C. Namely, green (A)  red (B) and blue (C), peaks indicate band edge excitons  for in-plane  and out-of-plane spin configurations of AFM aligned spins at +K and -K valleys,  respectively. Black arrows close to bands indicate spin degenerated bands. The position of the peaks are in order but not in scale. All of the presented direct transition are coupled to "z" polarized light with the intensities listed in Table \ref{tab:MPX3}. (i) The first BZ with labeled high symmetry k-points. }
\end{figure*}

\subsubsection*{Spin-orientation-dependent excitonic features in MnPX$_3$}

Interestingly, the excitonic properties are sensitive to the direction of the spins. Owing to the spin direction dependent effective masses observed in MnPX$_3$ presented in Fig. \ref{fig:vs} (d), we discuss below only these type of monolayers. The effect of the spin direction on E$_B$ is on the order of 1-2 meV in MnPS$_3$ but on the order of 10 meV in MnPSe$_3$ monolayer, as presented in Fig. \ref{fig:excitons} (b), irrespective of the value used for the Hubbard U parameter. Furthermore, opposite K-valleys (at $+$K and $-$K) show different effective masses and, consequently, different values of E$_B$, as shown in Fig. \ref{fig:excitons}(c). Hence, the E$_B$ depends on the orientation of the AFM aligned spins. Particularly, due to the sizeable valley-dependent curvature of the VBM (see Fig. \ref{fig:vs} (d)) observed for the MnPSe$_3$, the E$_B$ of excitons at $\pm K$ valleys differ by up to E$_B^{-}$ $-$ E$_B^{+}$ = 12 meV (for MnPSe$_3$ in vacuum).

Combining the information of the optical selection rules and exciton binding energies, we present in Fig.\ref{fig:transitions} the schematic diagram of optically active band edge transition considering the in-plane and out-of the plane direction of spins. Our calculations reveal that the emissions from the opposite $+K$ and $-K$ valleys are energetically below the onset of absorption spectrum. For the in-plane directions of the spins, the energy difference between $+K$ and $-K$ is rather small (separated about 1 meV and 0.1 meV for MnPSe$_3$ and MnPS$_3$, respectively, as shown in Figs. \ref{fig:transitions}(a,d)). On the other hand, for out-of-plane spins, the exciton emission from $\pm$K valleys are separated by 42 meV and 6 meV for MnPSe$_3$ and MnPS$_3$, respectively. The separation of the exciton peaks at $\pm K$ valleys (see difference between the position of the B and C peaks in Fig. \ref{fig:transitions}) is calculated as:
\begin{align}
\Delta_B & = \left[E_g(K-) - E_B(K-)\right] - \left[E_g(K+) - E_B(K+)\right] \nonumber \\
& =  \Delta E_B+\Delta^{VB}-\Delta^{CB} \, .
\end{align}
where $\Delta E_B = E_B(K-)-E_B(K+)$ is the difference between the exciton binding energies at $\mp K$ valleys and $E_g(K\pm)$ the electronic band gap at a given $K$ valley. We predict that one peak should be observed for in-plane oriented spins (green peak in Fig. \ref{fig:transitions}) whereas two peaks emerge for the spins deflected from the monolayer plane (red and blue peaks in Fig. \ref{fig:transitions}). Therefore, the energy $\Delta_B$ is a robust magnetic fingerprint of the AFM spin directions. Note that the energy separation of $\pm$K valleys depends on the dielectric environment, as presented in Fig. \ref{fig:excitons}(c), but should be visible in hBN encapsulated samples, specially in MnPSe$_3$ since the valley splitting is dominated by the electronic counterparts $\Delta^{CB}$ and $\Delta^{VB}$.

\section{Conclusions}
Here, we systematically  examine the MPX$_3$ materials
emphasizing  the role of spin reorientation, magnetic arrangement and electron correlation effects in various properties.  In this regard, the chemical trends in respect to the type of the chalcongen atom as well as transition metal are  examined. We have demonstrated that the electronic features such as band gaps, effective masses, dielectric screening and exciton binding energies  strongly depends on the type of chalcogen atom. Notably, larger effective masses are attained for S than Se contained monolayers (by about 30 $\%$), whereas smaller dielectric screening (up to $~40\%$) are reached for MPS$_3$. In general, the larger effective masses and smaller dielectric screening length are obtained for employed materials than for widely examined TMDs, resulting  in larger exciton binding energies of direct transitions of MPX$_3$ than corresponding monolayers of TMDcs. Generally, all examined band edge direct transitions of (Mn,Fe)PX$_3$ monolayers turned out to be optically active coupled to linear polarized light, with type of  polarization sensitive to  magnetic arrangement.

Regarding the direct band edge transitions, the MnPX$_3$ are of particular importance as could be exploited as promising valley electronics materials. In particular, we have shown that the valley splitting at the $\pm K$ can be effectively  controlled by the direction of the magnetic moments. In particular, the sizable valley splitting occurs for the out-of plane direction of the magnetic moments in monolayer of MnPSe$_3$ (smaller effect observed for MnPS$_3$), resulting in valley dependent gaps.  Namely, two distinct peaks are expected to be visible below the absorption onset for the out-of-plane AFM aligned spins, whereas one peak for the in-plane case. Hence, the number of the peaks and the separation of the peaks can be regarded as magnetic fingerprint of the orientation of AFM aligned spins. The change in the separation of the peaks points towards the plausible spin rotation. 

On the other hand,  the $\pm K$ valleys obtained for MnPX$_3$ are spin degenerated, which might hinder their spintronic applications. Hence, we have also propose a novel way how the  spin valley polarization can be attained. We have suggested that particular deformation of the hexagonal  lattice of MnPX$_3$ could lead to the spin resolved valley splitting. Our results give insight into the valley splitting realization in 2D antiferromagnets. In addition, the anistropic effective masses and the type of linear polarization can be regarded as magnetic markers probing the type of AFM arrangements. Finally, the spin dependent features  have  been identified such as valley splitting of VBM and CBM, the effective mass of holes and exciton binding energies. These features can be referred  as sensitive parameters that provide insight into spin flop transitions.


\section{Acknowledgment}

M.B. acknowledges financial support from the  University of Warsaw under the "Excellence Initiative - Research University" project. P.E.F.J. and J.F. acknowledge the financial support of the Deutsche Forschungsgemeinschaft (DFG, German Research Foundation) SFB 1277 (Project-ID 314695032, projects B07 and B11), SPP 2244 (Project No. 443416183), and of the European Union Horizon 2020 Research and Innovation Program under Contract No. 881603 (Graphene Flagship). T.W. acknowledges financial support of National Science Centre, Poland under grant no. 2021/41/N/ST3/04516. Access to computing facilities of the Interdisciplinary Center of Modeling (ICM), University of Warsaw are gratefully acknowledged. We gratefully acknowledge the Polish high-performance computing infrastructure PLGrid (HPC Centers: ACK Cyfronet AGH) for providing computer facilities and support within computational grants no. PLG/2022/015685 and PLG/2023/016571. We acknowledge ACK Cyfronet AGH (Poland) for awarding this project access to the LUMI supercomputer, owned by the EuroHPC Joint Undertaking, hosted by CSC (Finland) and the LUMI consortium through Pl-Grid organization (Poland), under the grant entitled: "Electronic, optical and thermoelectric properties of selected layered materials and selected heterostructures".

\pagebreak

\onecolumngrid
\begin{center}
  \textbf{\large Supplementary Material:\\"Magneto-optical anisotropies of 2D antiferromagnetic MPX$_3$ from first principles"}\\[.2cm]
  Miłosz Rybak,$^{3}$ Paulo E. Faria Junior,$^{2}$ Tomasz Woźniak,$^{3}$  Paweł Scharoch,$^{3}$  Jaroslav Fabian,$^{2}$ and Magdalena Birowska$^{1,*}$\\[.1cm]
  {\itshape ${}^1$University of Warsaw, Faculty of Physics, 00-092 Warsaw, Pasteura 5, Poland\\
  ${}^2$Institute for Theoretical Physics, University of Regensburg, 93040 Regensburg, Germany\\
  ${}^3$Department of Semiconductor Materials Engineering Faculty of Fundamental Problems of Technology Wrocław University of Science and Technology Wybrzeże Wyspiańskiego 27, 50-370 Wrocław, Poland\\}
  ${}^*$Electronic address: Magdalena.Birowska@fuw.edu.pl\\
\end{center}


\author{Magdalena Birowska}\affiliation{\IFT}


\maketitle
\renewcommand{\theequation}{S\arabic{equation}}
\renewcommand{\thefigure}{S\arabic{figure}}
\renewcommand{\thetable}{S\arabic{table}}
\renewcommand{\thesection}{S\arabic{section}}
\section*{Determination of  Hubbard U from \textit{Ab initio} }

\begin{figure}[!h]
    \vspace{5pt}
    \centering
    \includegraphics[scale=.5]{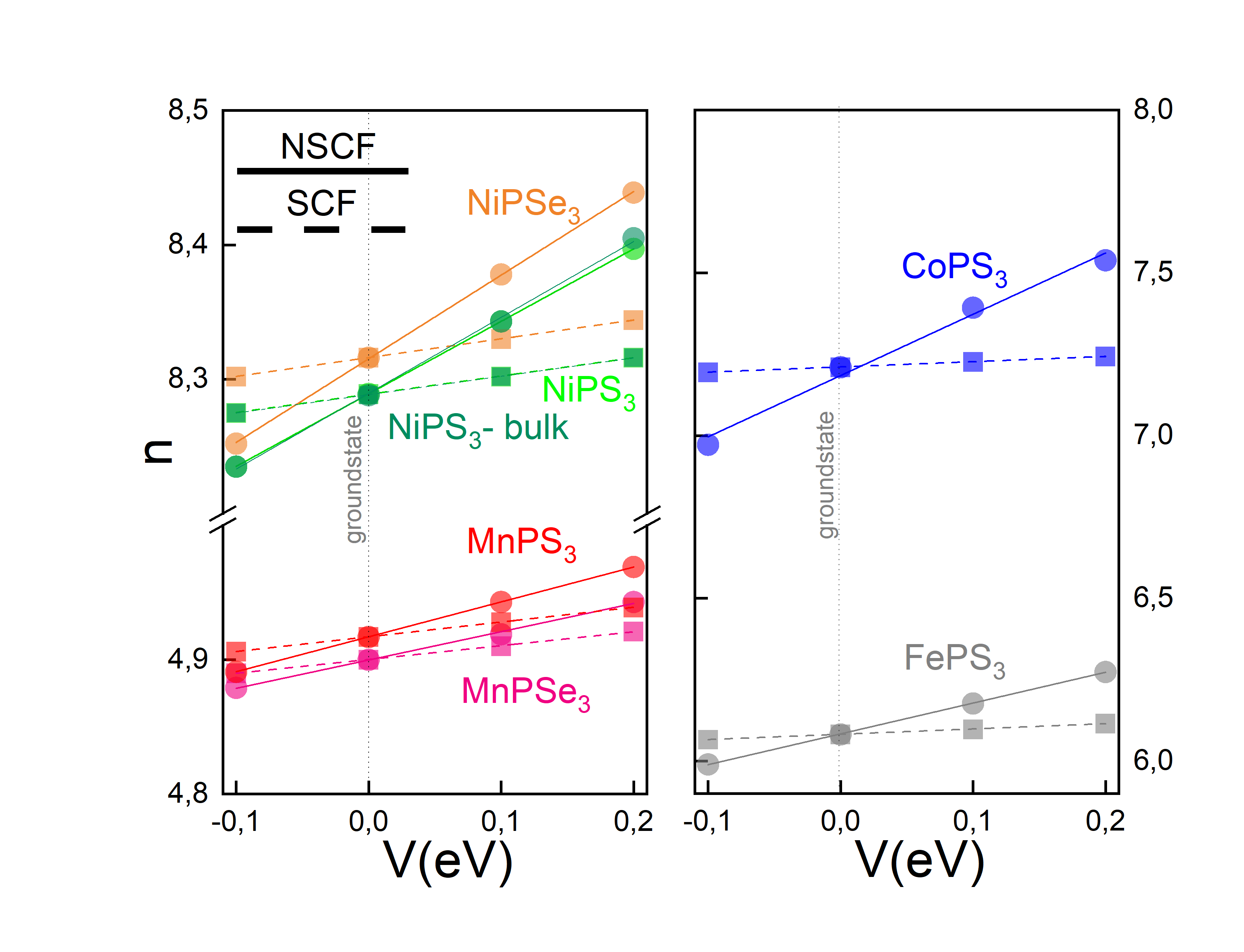}
    \vspace{0pt}
    \caption{Response functions  found from a linear fit of the number of d-electrons  as a function of the additional potential V:  (a) for $MPX_3 (M= Mn, Ni$, (b) for $M'PX_3 (M'= Co, Fe)$ }
\end{figure}
To estimate the Hubbard $U_{eff}$ parameter,  we adopt the linear response method proposed by method of Anisimov and co-workers in the framework of a
plane-wave pseudopotential approach implemented in VASP by Cococcioni \textit{et. al}  \cite{PhysRevB.71.035105}. The U parameter is  found from the linear fits for both  non-selfconsistent (NSCF) as and selfconsistent (SCF) response functions as defined $U\approx \left ( \frac{\partial N_{I}^{SCF}) }{\partial V_I} \right )^{-1}-\left ( \frac{\partial N_{I}^{NSCF}) }{\partial V_I} \right )^{-1}$. 
The results of U parameters for \textit{d} electrons of transition metal ions are collected in Table \ref{tab:U}.

\begin{table}[]
\centering
\caption{Calculated Hubbard $U_{eff}$ of \textit{d} electrons of transition metals for various MPX$_3$ materials. The values are given in  eV.}\label{tab:U}
\vspace{0.4cm}
  \def\arraystretch{1.5}
\begin{tabular}{c|c|c|c|c|c|c|c}
\hline
$MPX_3:$        & $MnPS_3$ & $MnPSe_3$ & $FePS_3$ & $NiPS_3$ & $NiPSe_3$ & $NiPS_3- bulk$ &$CoPS_3$\\
\hline
$U_{eff} [eV]: $ & 4.9      & 5.3       & 5.1      & 5.6      & 5.6       & 5.6  & 5.7  \\ 
\hline
\end{tabular}
\end{table}

\section*{Tables of collected electronic features.}
\begin{table*}[h!]
\caption{\label{tab:MPX3} Valence and conduction band edges (VBM, CBM) and electronic band gaps for various monolayer systems obtained for magnetic groundstate denoted as magn. state. The $\mu_m$ denotes the spin magnetic moment on metal atom (M= Mn, Ni, Fe).}  
 \def\arraystretch{1.5}
\begin{center}
\begin{tabular}{  c|  c| c| c|c |c|c }
   \hline
 ML of MPX$_3$ & U$_{eff}$ [eV] &magn. state& lattice constant [\AA{}]  & $\mu_m$ [$\mu_B$] &Band gap [eV] & VBM CBM  \\
 \hline
 MnPS$_3$&  5& AFM-N  & 6.108  &  4.610 & 2.502 (D)  & K   \\
         &  3& AFM-N  & 6.098  &  4.491 & 2.242 (D) & K \\
 \hline
 MnPSe$_3$  & 5& AFM-N & 6.424 & 4.606 &1.846 (D) &K\\
            & 3& AFM-N & 6.385 & 4.475 &1.662 &K\\
  \hline
 NiPS$_3$   &6.45 &AFM-zig& 5.856&1.621&2.288 &KY\\
            &5    &AFM-zig& 5.839&1.522&1.954 &KY\\
            &3    &AFM-zig& 5.819&1.378&1.572 &KY\\
  \hline
 NiPSe$_3$  &6.45 &AFM-zig & 6.184 & 1.566 & 1.883(D) & K\\
            &5    &AFM-zig & 6.168 & 1.457 & 1.6564   & KY \\
            &3    &AFM-zig & 6.146 & 1.291 & 1.258 & (0.282;0.5;0)Y\\
  \hline
 FePS$_3$   &5.3 &AFM-zig&5.988&3.704&2.423(D)&K\\
            &2.6   &AFM-zig&5.963&3.422&1.736  &$\Gamma$(0.273;0;0)\\
  \hline
 FePSe$_3$  &5.3& AFM-zig&6.304&3.687&1.825(D)&K\\
            &3  & AFM-zig&6.274&3.527&1.482(D)&K\\
  \hline
\end{tabular}
\end{center}
\end{table*}

\begin{figure*}[]
    \vspace{5pt}
    \centering
\includegraphics[width=1\textwidth]{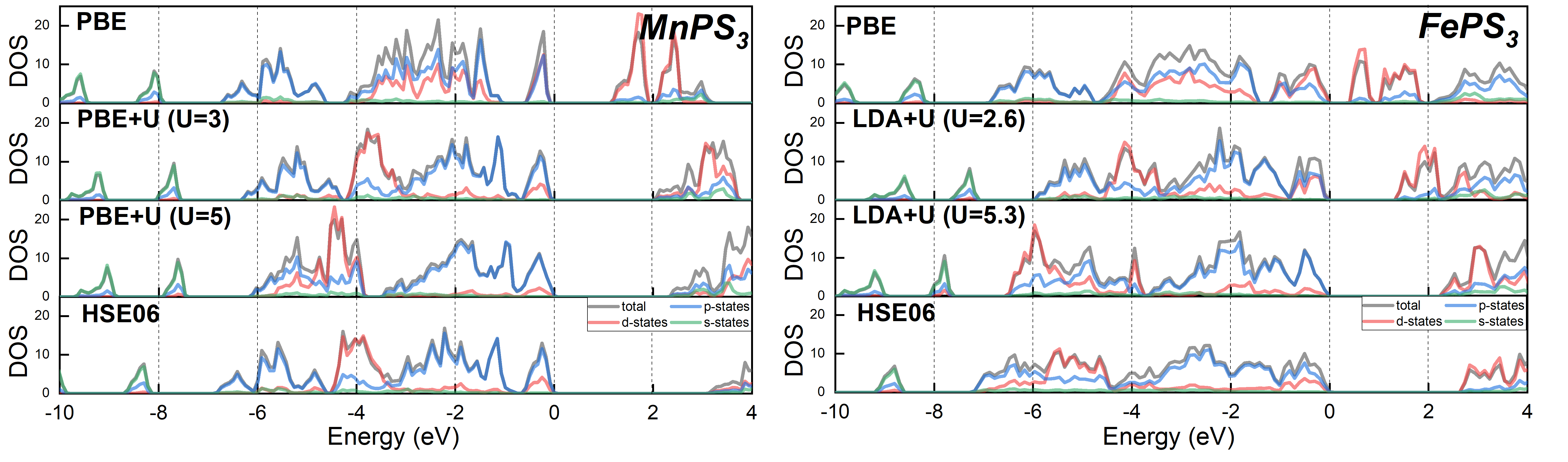}
    \vspace{0pt}
    \caption{Projected density of states (PDOS) assuming various exchange-correlation functional for MnPS$_3$ (right side) and FePS$_3$(left side).  }
\end{figure*}

\begin{figure*}[ht]
    \vspace{0pt}
    \centering
    \includegraphics[width=0.75\textwidth]{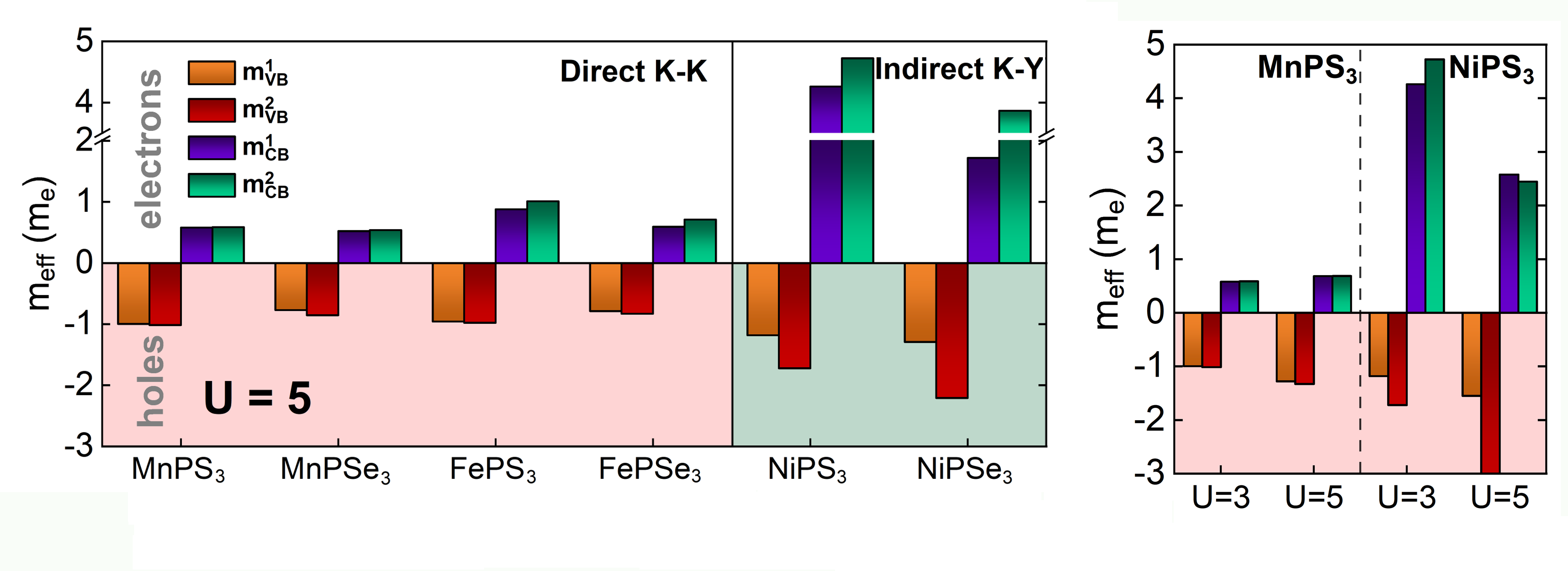}
    \vspace{0pt}
    \caption{In-plane components of the effective mass tensor for (left picture) all employed monolayers, (right picture) various Hubbard U.}
\end{figure*}

\begin{table*}[!h]
\caption{\label{tab:msses} In-plane components of the effective mass tensor ($m_1$, $m_2$). The $m_1$ and $m_2$ are given in $m_e$ unit and they are determined along principal axes (square brackets). The direction of the principal axes are given in  Cartesian coordinates in reciprocal space.}  
 \def\arraystretch{1.5}
\begin{center}
\begin{tabular}{  c|  c| c| c}
   \hline
 ML of MPX$_3$ & U${eff}$ [eV] & eff. mass of holes m$_1$ [princ. axis],  m$_2$ [princ. axis] & eff. mass of electrons m$_1$ [princ. axis],  m$_2$ [princ. axis]\\
 \hline
 MnPS$_3$&  5&    -0.957 [0.97 0.23 0]  -0.955 [-0.23 0.97 0] &0.577 [-0.39 0.93 -0.1] 0.588 [0.93 0.39 0.1] \\
    &    3&    -1.244 [0.97 -0.23 0] -1.246 [0.23 0.97 0] & 0.691 [0.99 -0.13 0] 0.689 [0.13 0.99 0]\\
 \hline
 MnPSe$_3$ &5& -0.767 [-0.39 0.93 0] -0.768 [0.93 0.39 0]   & 0.546 [0.93 -0.39 0] 0.546 [0.93 0.39 0]\\
    &3& -1.023 [0.97 -0.23 0] -1.026 [0.23 0.97 0] & 0.625 [ 0.95 -0.32 0] 0.623 [0.32 0.95 0]\\
     \hline

  NiPS$_3$   &5   & -1.183 [0 1 0] -1.722 [1 0 0] & 4.724 [0 1 0] 4.262 [1 0 -0.01] \\    
     &3   & -1.550 [0 1 0] -16.592 [1 0 -0.1] & 2.577 [1 0 -0.1] 2.447 [0 1 0] \\
     \hline     
NiPSe$_3$         &5   & -1.293 [0 1 0] -2.209 [1 0 0] & 3.865 [0.1 1 0] 1.718 [1.0 -0.1 -0.1]\\
    &3   & -1.380 [1 0 0] -4.628 [0 1 -0.1] & 1.869 [0.1 1 0] 1.031 [1 -0.1 -0.1]\\
     \hline 
FePS$_3$ &5.3& -0.927 [0 1 0.1] -0.969 [1 0 0] & 1.053 [0.1 1 0] 0.938 [1 -0.1 0]\\
     \hline 
FePSe$_3$ &5.3& -0.772 [0 1 -0.1] -0.788 [1 0 0] & 0.578 [-0.39 0.93 -0.1] 0.577 [0.93 0.39 0.1]\\
&3  & -1.056 [-0.01 1 0] -1.334 [1 0.01 -0.01] & 1.087 [0.04 1 0] 1.045 [1 -0.04 0]\\
     \hline 
\end{tabular}
\end{center}
\end{table*}

\begin{table*}[h!]
\centering
\caption{MnPSe$_3$ with and without SOI included. Note that the collinear spins are assumed. The SO offset is the energy difference between the results obtained with SOC and without it.}
\vspace{0.4cm}
  \def\arraystretch{1.5}
\begin{tabular}{c|c|c|cccc|cccc|c}
  \hline
method  & band edge  &   \textit{k-point} &
  \multicolumn{4}{c}{holes: \textit{m$_1$} [principle axis]} &
  \multicolumn{4}{c}{\textit{m$_2$} [principle axis]} &
  \textit{SO offset {[}meV{]}} \\
  \hline
\textit{without SO} &
  \textit{VB} &
  \textit{K/K'} &
  \textbf{-0.767} &
 [-0.38 &  0.92 &  0.00]&   \textbf{-0.768} &
  [0.92 &
  0.38 &
  0.00] &
  - \\
with SOC: in-plane  spins   & VB & K    & \textbf{-0.752} & [0.52 & 0.86  & 0.00] & \textbf{-0.825} & [0.86 & -0.52 & 0.00] & 0  \\
 & VB & K'   & \textbf{-0.767} & [0.74 & -0.67 & 0.00 ]& \textbf{-0.807} & [0.67 & 0.74  & 0.00] & 0   \\
with SOC: out-of-plane & VB & K    & \textbf{-0.752} & [0.52 & 0.86  & 0.00] & \textbf{-0.825} &[0.86 & -0.52 & 0.00] & -17.9      \\
 & VB & K'   & \textbf{-0.767} & [0.74 & -0.67 & 0.00] & \textbf{-0.807} & [0.67 & 0.74  & 0.00] & 0.21     \\
   \hline
method & band edge  &   \textit{k-point} &
  \multicolumn{4}{c}{electrons: \textit{m$_1$} [principle axis]} &
  \multicolumn{4}{c}{\textit{m$_2$} [principle axis]} &
  \textit{SO offset {[}meV{]}} \\
    \hline
\textit{without SO}                        & CB & K/K' & \textbf{0.547}  & [0.92 & -0.38 & 0.00] & \textbf{0.546}  & [0.38 & 0.92  & 0.00] & -      \\
with SOC: in-plane  spins   & CB & K    & \textbf{0.551}  & [0.83 & -0.56 & 0.00] & \textbf{0.542}  & [0.56 & 0.83  & 0.00] & -6.39  \\
& CB & K'   & \textbf{0.551}  & [0.59 & 0.81  & 0.00] & \textbf{0.541}  & [0.81 & -0.59 & 0.00] & -6.39 \\
with SOC: out-of-plane spins & CB & K    & \textbf{0.551}  & [0.83 & -0.56 & 0.00] & \textbf{0.542}  & [0.56 & 0.83  & 0.00] &  -3.50   \\
& CB & K'   & \textbf{0.551}  & [0.59 & 0.81  & 0.00] & \textbf{0.541}  & [0.81 & -0.59 & 0.00] & -36.60  \\
  \hline
\end{tabular}
\end{table*}

\begin{table*}[]
\centering
\caption{ MAE is the energy difference between the AFM aligned  spins along out-of plane (along easy axis) and in-plane directions (along hard axis), given in meV per magnetic ion. Note, that for in-plane directions the electronic structure is identical as without inclusion of the SOC. The $\Delta_{in-out}^{VB, CB(K\pm)}$ denotes the difference between the VBM (CBM) at given K point. The $E_g^{in}$ ($E_g^{out}$) electronic band gap for the in-plane (out-of-plane) direction of the AFM aligned spins. } 
\vspace{0.4cm}
  \def\arraystretch{1.5}
\begin{tabular}{c|c|c|c|c|c|c|c|c}
\hline
 & U [eV] & MAE   & $\Delta_{in-out}^{VB(K+)}$ [meV] & $\Delta_{in-out}^{CB(K+)}$ [meV]& $\Delta_{in-out}^{VB(K-)}$[meV] & $\Delta_{in-out}^{CB(K-)}$ [meV] & $E_g^{in}$ [meV] & $E_g^{out}$ [meV] \\
 \hline
MnPS$_3$ & 3 & 0.041 & 0 & 0 & 5.5  & 9.3 & 2.203  & 2.194   \\

& 5 & 0.028 & 0 & 0   & 6.5 & 3.9  & 2.483  & 2.479   \\

 & 7 & -  & 0  & 0   & 6.9 & 1.9  & 2.667  & 2.665   \\
 \hline
MnPSe$_3$ & 3 & 0.315 & 0        & 34.4     & 17.8      & 0         & 1.689  & 1.654   \\

& 5 & 0.223 & 0        & 27.9     & 15.4      & 0         & 1.853  & 1.825   \\

 & 7 & -  & 0  & 16.4     & 12.6  & 0   & 1.960  & 1.944  \\
 \hline
\end{tabular}
\end{table*}

\begin{table*}[]
\centering
\caption{ Electronic band edge features such as valley splitting ($\Delta^{VB}$=E$^{VB}_{K_-}-$E$^{VB}_{K+}$, $\Delta^{CB}$=E$^{CB}_{K_-}-$E$^{CB}_{K+}$), and spin degeneracy ($\delta^{VB}$=E$^{VB}_{K_-}-$E$^{VB}_{K+}$, $\delta^{CB}$=E$_{B}^{K_-}-$E$_{B}^{K+}$,  for monolayers of MnPX$_3$ and FePX$_3$, respectively for in-plane and out-of-plane directions of AFM aligned spins. } 
\vspace{0.4cm}
  \def\arraystretch{1.5}
\begin{tabular}{l|ll|ll|ll|ll}
\hline
         valley splitting  & MnPSe$_3$ & for U=3eV    & MnPSe$_3$ & for U=5eV  & MnPS$_3$  & for U=3eV   & MnPS$_3$  &for  U=5eV    \\
         \hline
 & $\Delta^{VB}$     & $\Delta^{CB}$     & $\Delta^{VB}$      & $\Delta^{CB}$   & $\Delta^{VB}$ & $\Delta^{CB}$   & $\Delta^{VB}$  & $\Delta^{CB}$  \\
in-plane spins   & -5.7    & 0     & 0      & 0    & 0      & 0    & 0      & 0     \\
out-of-plane spins & -21.5   & 38.6  & -17.7   & 33.1 & -6.6    & -10.3 & -6.4    & -4.4   \\
    \hline
   Spin degeneracy of the band edges         & FePS$_3$  & for U=2.6eV & FePS$_3$  & for U=5.3eV & FePSe$_3$ & for U=3eV & FePSe$_3$ & for U=5.3eV \\
  $\delta$          & $\delta^{VB}$     & $\delta^{CB}$     & $\delta^{VB}$      & $\delta^{CB}$   & $\delta^{VB}$      & $\delta^{CB}$   & $\delta^{VB}$      & $\delta^{CB}$    \\
in-plane spins   & 0      & 7.5   & 1.8    & 9.1  & 7.4    & 5.5  & 7.4    & 3.1   \\
out-of-plane spins & 0      & 6.4   & 1.5    & 7.4  & 11.5   & 8.2  & 9.3    & 5   \\
\hline
\end{tabular}
\end{table*}

\begin{table}[h]\footnotesize
\caption{ Static dielectric constants $\varepsilon_{ij}$ (electronic contribution) and 2D polarizability $\chi^{\parallel}$ calculated for the monolayer systems and U=3 eV. The static dielectric properties are calculated by means of density functional perturbation theory  in the independent particle approach (IP) neglecting local field effects and including local fields effect (TD-DFT)  } \label{tab:dielectric}
 \def\arraystretch{1.7}
\begin{center}
\begin{tabular}{  c|  c|  c|  c| c }
   \hline
Magn. & U [eV] & method & $\varepsilon_{xx}$ $\varepsilon_{yy}$, & $\chi^{\parallel}_{xx}$ $\chi^{\parallel}_{yy}$ [\textrm{\AA}] \\
   \hline
MnPS3 & 5 & IP& 2.35 2.35 &  3.25 3.25  \\
      & 5 & TD-DFT& 2.16 2.16  &2.80 2.80 \\ 
  \hline
MnPSe3 & 5 & IP& 2.60 2.60 & 4.50 4.50    \\
      & 5 & TD-DFT& 2.42 2.42 & 3.99 3.99 \\ 
  \hline
NiPS3 & 5 & IP& 2.84 2.82  & 4.28 4.23\\
      & 5 & TD-DFT&  2.70 2.68 & 3.94 3.90  \\
  \hline
NiPSe3 & 5 & IP& 3.31 3.26  &  6.13 6.00  \\
      & 5& TD-DFT& 3.18 3.13  & 5.76 5.65 \\ 
  \hline
FePS3 & 5.3 & IP& 3.20 3.21 & 3.51 3.52  \\
      & 5.3 & TD-DFT& 2.96 2.95  & 3.13 3.10 \\ 
  \hline
FePSe3 & 5.3 & IP& 2.81 2.78 & 4.90 4.84   \\
      & 5.3 & TD-DFT& 2.63 2.61  & 4.43 4.38 \\ 
  \hline        
\end{tabular}
\end{center}
\end{table}

\begin{table}[h]\footnotesize
\caption{ Static dielectric constants $\varepsilon_{ij}$ (electronic contribution) and 2D polarizability $\chi^{\parallel}$ calculated for the monolayer systems and Hubbard U=3eV. The static dielectric properties are calculated by means of density functional perturbation theory  in the independent particle approach (IP) neglecting local field effects and including local fields effect (TD-DFT)  } \label{tab:dielectric}
 \def\arraystretch{1.7}
\begin{center}
\begin{tabular}{  c|  c|  c|  c| c }
   \hline
Magn. & U [eV] & method & $\varepsilon_{xx}$ $\varepsilon_{yy}$, & $\chi^{\parallel}_{xx}$ $\chi^{\parallel}_{yy}$ [\textrm{\AA}] \\
   \hline
MnPS3 & 3 & IP& 2.44 2.44  & 3.45 3.45  \\
      & 3 & TD-DFT& 2.26 2.26 & 3.02 3.02 \\ 
  \hline
MnPSe3 & 3 & IP& 3.02 3.02 & 4.82 4.82  \\
      & 3 & TD-DFT& 2.82 2.82 & 4.33 4.33 \\ 
  \hline
NiPS3 & 3 & IP& 3.15 3.10 & 5.12 5.01\\
      & 3 & TD-DFT& 3.02 2.98  & 4.82 4.72  \\
  \hline
NiPSe3 & 3 & IP& 4.14 4.02  & 7.51 7.21  \\
      & 3& TD-DFT& 3.02 2.98 & 4.82 4.72 \\ 
  \hline
FePSe3 & 3& IP& 3.06 3.03 &  5.56 5.49  \\
      & 3 & TD-DFT& 2.91 2.89  & 5.15 5.09 \\ 
      \hline
\end{tabular}
\end{center}
\end{table}

\begin{table}[h]\footnotesize
\caption{ Exciton binding energies at different K valleys and environments (vacuum, hBN) as a function of spin direction $\Theta$ for  MnPX$_3$ (X=S, Se). } \label{tab:dielectric}
 \def\arraystretch{1.8}
\begin{center}
\begin{tabular}{  c| c| c|  c|  c| c |c }
   \hline
system & U [eV]& spin direction $\Theta$ [$^0$] & E$_b^{K+}$ in vacuum & E$_b^{K+}$ in hBN & E$_b^{K-}$ in vacuum  & E$_b^{K-}$ in hBN  \\
   \hline
MnPS$_3$ & U=3  & 0  &  1032.19 & 344.46 & 1034.46 & 346.02 \\
& & 50 &  1032.47 & 344.67  & 1034.39 & 345.83  \\
& & 90  & 1033.37 & 345.34 & 1033.48 & 345.45  \\
  \hline
 & U=5  & 0   & 1023.98 &  322.49& 1025.60 &323.21 \\
& & 50 &  1023.98&322.49 & 1025.60 &323.21 \\
&  &90 &  1024.69& 322.79 &1024.54 & 322.85 \\
  \hline
 MnPSe$_3$ & U=3 &   0   & 770.25 &266.92 &787.82& 278.42 \\
  && 50  & 773.76 &269.31 &785.75& 276.89 \\
 & &90  & 780.62& 273.32 &779.96& 273.13\\
  \hline
 & U=5& 0  &  773.71 &254.23& 785.49& 261.16\\
& &  50 &  775.00& 254.61& 781.97 &259.01\\
&  & 90  & 778.86& 256.97 &779.77& 257.45 \\
  \hline
\end{tabular}
\end{center}
\end{table}

\begin{table}[h]\footnotesize
\caption{ Exciton binding energies for MLs considering different values of the effective dielectric constant of
the surroundings  $\varepsilon$, calculated using  two approaches (IP and TD-DFT) for dielectric screening length of ML.  } \label{tab:dielectric}
 \def\arraystretch{0.9}
\begin{center}
\begin{tabular}{  c| c| c|  c|  c|c  }
   \hline
ML& $\varepsilon$   &   U=3 eV, IP   &   U=3 eV, TD-DFT  &    U=5 eV, IP  &    U=5 eV, TD-DFT \\
\hline
 & 1.0  &   1028.58 &  1132.13    &1019.06  & 1132.42\\
 & 1.5  &    814.46 &   890.75   &   797.34  & 879.03\\
 & 2.0  &    673.53 &   732.35   &  652.41  &  714.90\\
 & 2.5   &   571.60  & 618.59   &   549.11  &  598.30\\
MnPS$_3$  & 3.0   &   494.27  &  532.73  &     470.56 &   510.31\\
 & 3.5   &   433.05  &  464.87  &    409.35 & 442.17\\
 & 4.0   &   383.39 &   410.52   &    360.20 &   387.07\\
 & 4.5  &   342.54  & 365.24   &  319.61  & 342.01\\
 & 5.0   &   308.07 &   327.64  &    285.84 &   304.97\\
  \hline
  &1.0   &  778.82  &  843.61   &   778.93  &  850.94\\
  &1.5   &  623.37 &   671.57  &    616.20 &   668.73\\
  &2.0   &  519.65  &  557.50  &    508.78  &  549.51\\
 & 2.5   &  444.61 &   475.08  &    431.41  &  463.95\\
MnPSe$_3$ & 3.0    & 386.72 &   412.18  & 372.29&   398.80\\
 & 3.5   &  340.69 &   362.07   &   325.94  &  347.92\\
 & 4.0  &   303.63 &   321.54   &   288.12  &  306.50\\
&  4.5   &  272.23  &  287.72   &   256.96  &  272.58\\
 & 5.0  &   246.07&   259.44   &   230.95   & 244.25\\
 \hline
 & 1.0  &   985.67 &  1031.65 &  1073.15 &  1140.95\\
  &1.5  &   825.85 &  862.46  &  886.32  &  939.24\\
 & 2.0  &   716.31 &  746.68  &  759.50  &  802.73\\
 & 2.5  &   634.46 &  660.29  & 665.62   & 701.86\\
NiPS$_3$ & 3.0  &   569.94 &  592.46  &  592.25  &  623.30\\
 & 3.5  &   517.45 &  537.29  &  532.90  &  559.77\\
 & 4.0  &   473.66 &  491.15  &  483.65  &  507.35\\
 & 4.5  &   436.31 &  451.94  &  442.10  &  462.95\\
 & 5.0  &   404.01 &  418.21  &  406.42  &  424.93\\
  \hline
 & 1.0  &  670.36 &  940.83 &      805.38&   844.93\\
 & 1.5  &  562.14 &  775.12 &      673.12 &  704.36\\
 & 2.0 &   487.23  & 662.72 &      582.05 &  608.00\\
 & 2.5  &  430.83 &  579.42 &      513.95 &  536.08\\
NiPSe$_3$ & 3.0 &   386.65 &  514.41 &      460.50 &  479.60\\
&  3.5  &  350.42 &  462.08 &      416.84 &  433.64\\
 & 4.0  &  320.06 &  418.51 &      380.60 &  395.41\\
&  4.5  &  294.47&   381.96  &     349.65 &  362.93\\
&  5.0 &  272.19 &  350.44  &     323.07  & 334.86\\
  \hline
&  1.0   &  1038.18 &  1132.84 &-&-\\
&  1.5    & 825.91 &   896.20&-&-\\
&  2.0    & 685.68 &   740.31&-&-\\
&  2.5    & 584.24 &   628.15&-&-\\
FePS$_3$ &  3.0   &  506.71 &  543.07&-&-\\
&  3.5   &  445.41 & 475.55&-&-\\
&  4.0   &  395.42 & 420.84&-&-\\
&  4.5   &  354.01 & 375.72&-&-\\
&  5.0   &  319.04 & 338.00  &-&-\\
\hline
&  1.0  &  766.45&  812.40 &    741.87&    798.61\\
 & 1.5 &   626.08&  661.47 &    590.19  &  632.78\\
 & 2.0 &   530.94 & 559.68 &    489.66 &   522.38\\
&  2.5    &461.05&  484.65 &    416.52 &   443.24\\
FePSe$_3$ &  3.0 &   406.60&  426.59  &360.88 &  382.75\\
 & 3.5 &   362.81 & 379.87 &    316.85 &   334.78\\
&  4.0  &  326.70 & 341.66  &   281.01  &  296.30\\
 & 4.5 &   296.51 & 309.55 &    251.24  &  264.49\\
&  5.0  &  270.58 &282.16  &   226.68   & 237.58\\
  \hline
  \end{tabular}
\end{center}
\end{table}

\begin{figure*}[ht]
    \centering
\includegraphics[width=0.8\textwidth]{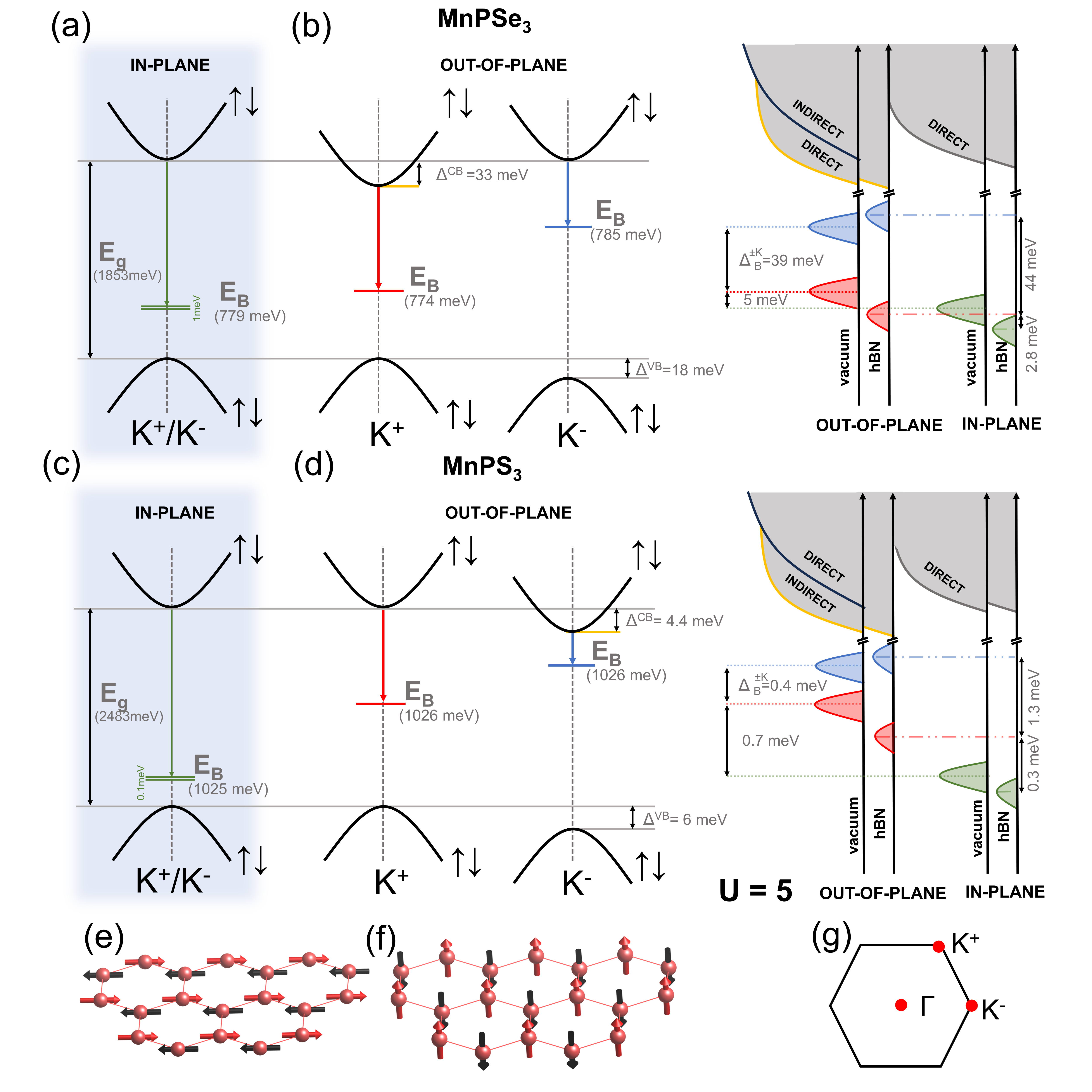}
    \vspace{0pt}    \caption{\label{fig:transitions} Schematic diagrams (U=5 eV) of   direct transitions for (a, b) MnPSe$_3$ and (c, d) MnPS$_3$ exhibiting in-plane and out-of-plane directions of the AFM aligned spins, as schematically denoted  at the bottom of the picture (e, f). On the right side of the pictures schematic absorption onsets with depicted in coloured exciton peaks. Namely, blue, red and green peaks indicate band edge excitons  for out-of-plane spin configurations at +K and -K valleys, and in-plane configuration of AFM aligned spins, respectively.   Black arrows close to bands indicate spin degenerated bands. The position of the peaks are in order but not in scale. All of the presented direct transition are coupled to "z" polarized light with the intensities listed in Table \ref{tab:MPX3}. (g) The first BZ with labeled high symmetry k-points. }
\end{figure*}

\clearpage
\bibliography{references}

\end{document}